\documentclass[prl,twocolumn,aps,amssymb,footinbib,showpacs]{revtex4-1}
\usepackage{amssymb}
\usepackage{graphicx}
\usepackage{amsmath}
\usepackage{color}
\usepackage{subfigure}
\usepackage{setspace}
\usepackage{bm}

\newcommand {\beq} {\begin{equation}}
\newcommand {\eeq} {\end{equation}}
\newcommand {\bqa} {\begin{eqnarray}}
\newcommand {\eqa} {\end{eqnarray}}
\newcommand {\ca} {\ensuremath{c^\dagger}}

\newcommand {\up} {\ensuremath{\uparrow}}
\newcommand {\dn} {\ensuremath{\downarrow}}

\newcommand{\nbr} {\ensuremath{\langle ij \rangle}}

\begin{document}

\begin{abstract}
We study the finite temperature antiferromagnetic phase of the ionic
Hubbard model in the strongly interacting limit using quantum Monte Carlo based dynamical mean field theory. We find that the ionic potential plays a dual role
in determining the antiferromagnetic order. A small ionic potential
(compared to Hubbard repulsion) increases the super-exchange coupling
in the projected sector of the model, leading to an increase in the
 Neel temperature of the system. A large ionic potential leads to
resonance between projected antiferromagnetically ordered
configurations and density ordered configurations with double
occupancies, thereby killing antiferromagnetism in the system. This
novel way of degrading antiferromagnetism leads to spin polarization
of the low energy single particle density of states. The dynamic
response of the system thus mimics ferromagnetic behaviour, although 
the system is still an antiferromagnet in terms of the static spin
order.
\end{abstract}

\title{ Ferromagnetic response of a ``high-temperature'' quantum antiferromagnet}
\author{Xin Wang$^1$, Rajdeep Sensarma$^2$ and Sankar Das Sarma$^1$}
\affiliation{$^1$Condensed Matter Theory Center, Department of Physics, University of Maryland, College Park, Maryland 20742, USA\\
$^2$Department of Theoretical Physics, Tata Institute of Fundamental Research, Mumbai 400005, India}
\pacs{75.10.-b, 67.85.-d, 72.25.-b}

\maketitle

Ultracold atoms in optical lattices~\cite{BlochReview} have emerged as
a novel platform for strongly correlated physics, where lattice
models, relevant to condensed matter systems and other arenas of
physics, can be implemented and studied in a controllable
way~\cite{Greiner02,Esslinger08,Bloch08,Greiner10}. The easy
tunability of the Hamiltonian parameters and the accurate knowledge
and control of the underlying parameters have made these systems the
foremost candidate for emulating models like the repulsive Bose and
Fermi Hubbard model, which are routinely used as paradigms in
the study of correlation-driven superfluid-insulator
transitions~\cite{FisherBH89} and high temperature
superconductors~\cite{AndersonSc87}. In fact, a whole new
  subject at the interface of condensed matter and atomic physics has
  emerged in this context, being dubbed ``Optical Lattice Emulation''
  (OLE).

  The implementation of the Fermi Hubbard
  model~\cite{Esslinger08,Bloch08} in the strongly interacting limit
  has raised the prospect of observing antiferromagnetic (AF) spin
  ordering in these systems~\cite{Esslinger08}. At present, observing
  antiferromagnetism with cold atoms is a major goal of
  experimentalists, which would be a stepping stone towards
  observation of the more complicated phenomenon of high temperature
  superconductivity. While the basic mechanism of super-exchange has
  been verified in cold atom experiments~\cite{BlochSuper-Exchange},
  spontaneous AF ordering is yet to be seen in optical lattices. The
  main problem is that the temperature scale for the AF transition is
  simply too low to be achievable in the laboratory at the present
  time~\cite{AFTEMP}. While a lot of effort has been spent towards
  improving cooling techniques in optical lattices
  \cite{COOLING} , in this Letter, we propose a different way of
  observing AF ordering, by modifying the lattice model in a
  way that the Neel temperature is higher than that of the standard
  Fermi Hubbard model. A similar approach has been used to study
  ``magnetism'' in a system with an effective electric field on the
  atoms~\cite{GreinerEfield}. However, the magnetism in that case is
  driven by the hopping of fermions~\cite{SenguptaEfield} and throws
  little light on super-exchange dominated AF order, which is of
    key interest in the context of strong correlation physics.

We study a related model called the ionic Hubbard model,
which was first introduced in the context of ionic to neutral
transition of charge transfer organic compounds~\cite{IonHubbardearly,NagaosaJPSJ86}.
In addition to hopping of the fermions on a bipartite
lattice and the usual on-site Hubbard repulsion,
the fermions on the different sublattices feel different local potential
energies, which breaks the sublattice symmetry in the ``charge''
sector. This model can be easily implemented in a cold
atom setting, with the staggered local potential being imprinted using
holographic techniques~\cite{GreinerMicroscope}. In the strongly interacting limit, we find that the effective
super-exchange scale, and hence the AF transition temperature at
half-filling, increases with the ionic potential. This should make it
easier to see AF order in this model as compared with the standard
fermionic Hubbard model. We find that the Neel temperature can be
enhanced by about $40 \%$  for reasonable values of the parameters.

\begin{figure}[t]
\includegraphics[width=\columnwidth]{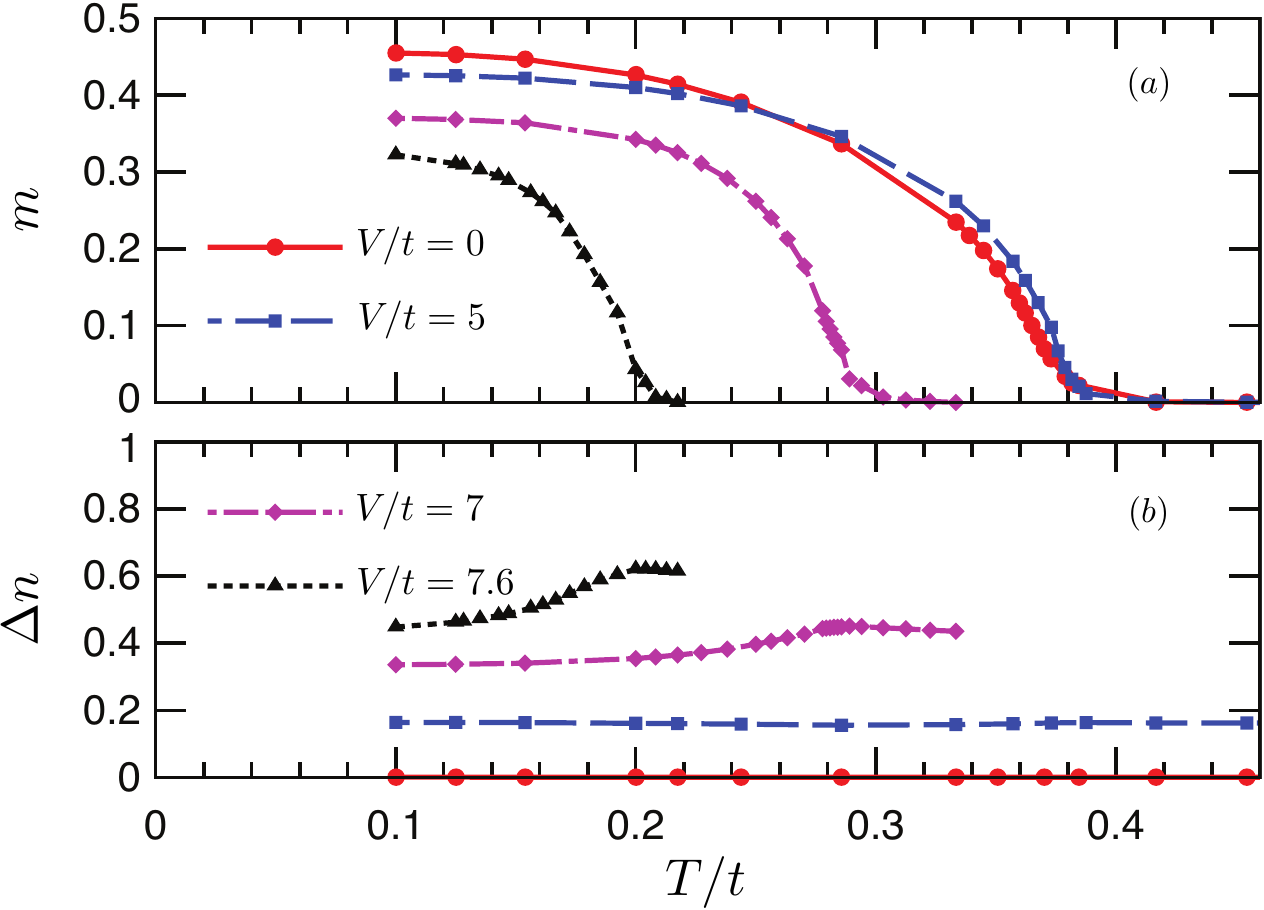}
\caption{(a) The staggered magnetization, $m$ and (b) the difference
  in density between $A$ and $B$ sublattices, $\Delta n$ as functions
of temperature for different values of the ionic potential $V$. The
Hubbard interaction is fixed at $U/t=10$ for all the plots.}
\label{fig:1}
\end{figure}   

In the ionic
Hubbard model, when the staggered potential is comparable to the
Hubbard repulsion, we find that the low energy single-particle density of states  (DOS)  for
the up $(\up)$ and down $(\dn)$ spins are different, i.e. the system shows
ferromagnetic characteristics in its low frequency dynamics, although
it continues to exhibit static AF order. This
counter-intuitive result is understood as an effect of the interplay
between the bichromatic nature of the lattice and the
AF order in the system.  This surprising result 
can be
confirmed by either polarization dependent rf spectroscopy~\cite{CARFSpec} or spin
conductivity~\cite{CASpincond} or spin injection spectroscopy~\cite{CASpinInj} measurements.

The ionic Hubbard model is defined on a bipartite lattice (e.g. a
cubic lattice with two
sublattices A and B) as 
\beq
H =-t\sum_{\nbr}\ca_{i\sigma}c_{j\sigma}+U\sum_i
n_{i\up}n_{i\dn}+\frac{V}{2}\sum_i (-1)^{\gamma_i} n_i,
\eeq
where $t$ is the nearest neighbour hopping matrix element, $U$ the
local Hubbard repulsion, $\gamma_i=1(0)$ if $i$ is a site on A (B)
sublattice, and $V$ is the amplitude of the staggered ionic potential.
This model has a rich phase diagram as temperature, carrier density,
the interaction parameter $U/t$ or the 
ionic potential $V/U$ are tuned~\cite{IonHubbardearly,NagaosaJPSJ86,Martin01,LopezSancho05,GargDMFT,Scalettar07,Dagotto07,Aligia04}. In this
Letter, we solely focus on the system at half-filling, i.e. one
particle per lattice site. In the non-interacting limit ($U=0$), the
system is a band insulator (since the staggered potential doubles the
unit cell). In the strongly interacting limit ($U\gg V, t$), the
system is an AF Mott insulator, with the spin ordering governed by a
super-exchange scale $J=4t^2/U$. Most of the previous
studies~\cite{IonHubbardearly,NagaosaJPSJ86,Martin01,LopezSancho05,GargDMFT,Scalettar07,Dagotto07,Aligia04}
on the system have focused on how the system goes from a band to a
Mott insulator and whether there is an intervening metallic phase,
while the antiferromagnetic phase has been studied at weak
coupling~\cite{Byczuk09, Garg13}. Our
work has a completely different focus. Starting from the interaction
dominated limit ($V=0$, $U/t \gg 1$), we are interested in the fate of
the spin ordering in the system as a function of the temperature and
the staggered potential $V/U$.
\begin{figure}
\includegraphics[width=0.9\columnwidth]{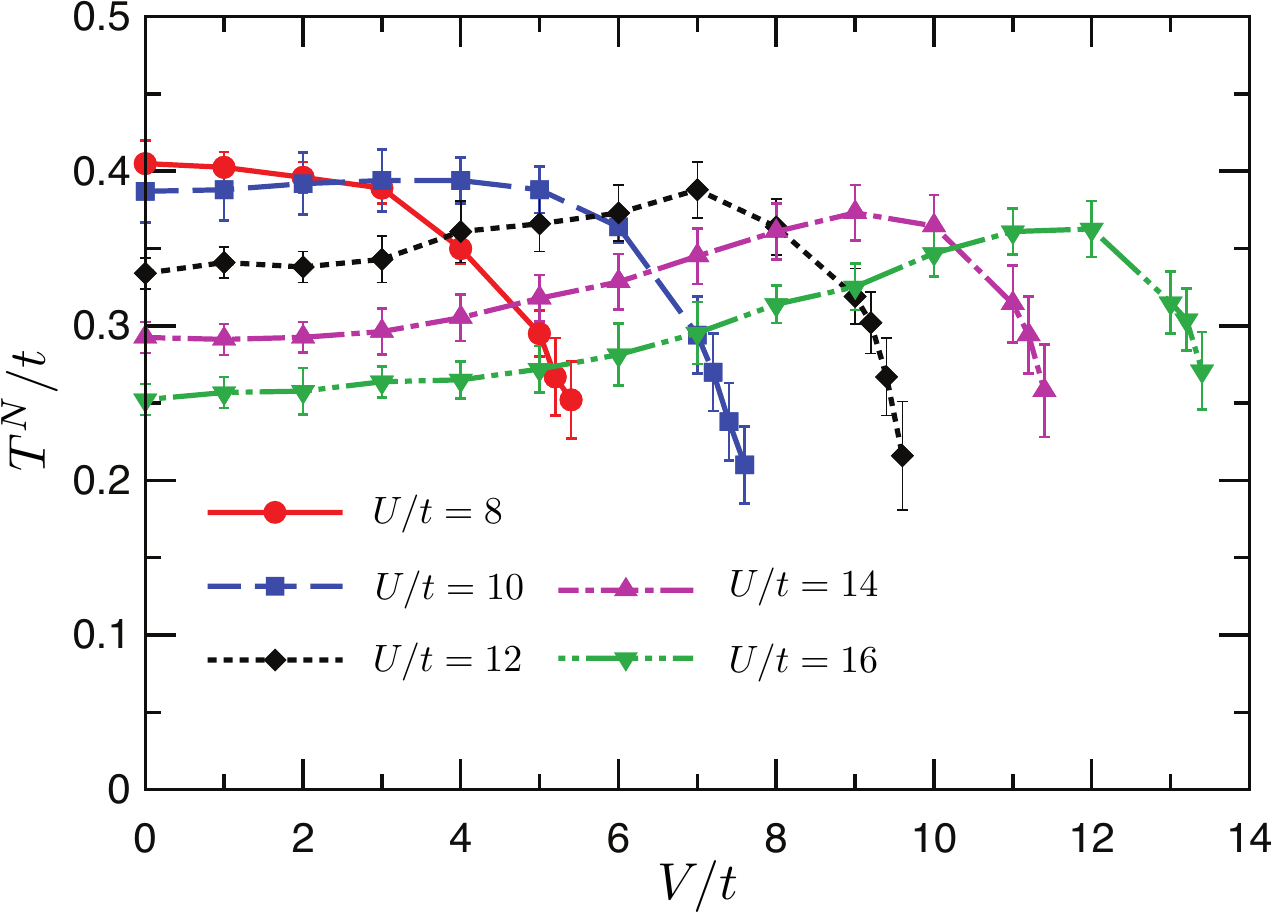}
\caption{The dependence of the Neel temperature $T^N$ on the ionic potential
$V/t$ for several different values of the Hubbard interaction
$U$. At large $U/t$, the Neel temperature initially rises with $V/t$
before crashing down.}
\label{fig:2}
\end{figure}

We use the dynamical mean field theory (DMFT)~\cite{Georges96}
together with CT-HYB quantum Monte Carlo impurity solver~\cite{Gull10}
to study the AF ordered phase of this model.  DMFT approximates the
interacting lattice problem by a single site or a small cluster
(i.e. impurities) interacting with a bath. The dynamics of the bath
and the impurities are then solved self-consistently to obtain the
local Green's functions for the interacting problem, which are used to
calculate various properties of the system. Since we are interested in
the spin dynamics in the presence of an explicitly broken sublattice
symmetry, the local Green's functions include
$G^{\up(\dn)}_{A(B)}(\tau)$, where $G^\sigma_\alpha(\tau)$ is the
imaginary-time Green's function of the fermions with spin $\sigma$ on
sublattice $\alpha$. We first obtain an AF state at $V=0$, after which
we slowly turn on $V$. We typically take at least $10^9$ Monte Carlo
steps in each iteration, and, for certain cases, more than 60
iterations are used to ensure convergence.  We note that as a
method, DMFT is exact in infinite dimensions. For hypercubic
lattices, DMFT calculations are expected to better capture the
qualitative and quantitative features of the model \cite{Georges96} as the
dimension increases, which, for cold atom systems, can at best
take the value of 3.

  We first focus our attention on the AF ordering in the
  system.  The density of each spin on each sublattice is given by 
    $n^\sigma_\alpha=1+G^\sigma_\alpha(\tau\rightarrow0^+)$, while the
  staggered magnetization, $m$, characterizing the AF order, is given
  by $m=(n^\up_A+n^\dn_B-n^\dn_A-n^\up_B)/4$. The dependence of $m$ on
  temperature for different values of $V/t$ (for a fixed $U/t=10$) is
  plotted in Fig.~\ref{fig:1}(a). The magnetization decreases with temperature and vanishes at the Neel temperature
  $T^N$. 
As $V/t$ is raised to around $7$ and beyond, the
  magnetization (at a given temperature) is sharply suppressed with
  increasing $V/t$, and $T^N$ also varies
  rapidly with $V/t$ in this regime.

 The Neel temperature is plotted as a function of $V/t$
for several values of $U/t$ in Fig.~\ref{fig:2}. Upto $U/t=10$, $T^N$ is a monotonically decreasing function of $V/t$. For
 $U/t=12$, $T^N$ first rises as a function of $V/t$,
reaches a maximum, and then crashes as $V/t$ is increased
further. For $U/t >12$ , this effect is much more pronounced.
This non-monotonic behaviour of $T^N$ at strong coupling can
be understood by the following perturbative argument. For large $U/t$
and small $V/U$, the Hubbard repulsion is the largest scale in the
problem and hence states with double occupancies are projected out of
the  low energy sector.
A Schrieffer-Wolff
type canonical transformation, $e^{-iS}$~\cite{GirvineiS} can then be used to
perturbatively include effects of virtual transitions to the high
energy sector (with finite double occupancies). The transformation is
obtained by assuming that the transformed Hamiltonian
$\tilde{H}=e^{iS}He^{-iS}$ does not have any term connecting low
and high energy sectors. To first order in $t/U$, 
\beq
iS= \sum_l \frac{T^1_{AB}(l)-T^{-1}_{BA}(l)}{U+V}+\frac{T^1_{BA}(l)-T^{-1}_{AB}(l)}{U-V}
\label{eqn:is}
\eeq
where $T^\eta_{AB}(l)$ is the part of the hopping operator on the bond
$l$ which hops a fermion from $A$ to $B$ sublattice. 
and increases the double occupancy of the system by $\eta =0,1$, or
$-1$. At half-filling, the effective low energy Hamiltonian is given by 
\beq
\tilde{H}= \frac{J}{1-\frac{V^2}{U^2}}\sum_{\nbr}\left( \vec{S}_i\cdot \vec{S}_j-\frac{1}{4}n_in_j\right)
\label{eqn:Heisenberg}
\eeq
where the spin operator
$\vec{S}_i=\ca_{i\sigma}\vec{\sigma}_{\sigma\sigma^{'}}c_{i\sigma^{'}}$,
and $J=4t^2/U$~\cite{NagaosaJPSJ86} is the standard Heisenberg
super-exchange scale. Physically, there are two possible processes
leading to the spin-spin interaction, where the double occupancy in
the intermediate virtual state is formed on the $B$ or the $A$
sublattice respectively. They contribute with the scale $\sim
t^2/(U\pm V)$ respectively, leading to an enhancement of the
super-exchange scale for
small $V/U$.  This can be contrasted with the extended Hubbard model
with nearest neighbour interaction (which does not break sublattice
symmetry), where the effective super-exchange coupling increases with
the nearest neighbour interaction $V'$ as
$\sim t^2/(U-V')$~\cite{Ruckenstein87,Dongen94}. In the strongly interacting limit of the Hubbard
model ($V=0$) at half-filling, the Neel temperature scales with the
Heisenberg coupling $J$, with numerical simulations yielding $T_N/J
\sim 0.957$ on the cubic lattice~\cite{Staudt00}. The same scaling
should hold in the small $V/U$ limit, where the low energy subspace
does not contain any configuration with double occupancy, 
  explaining the increase of $T^N$ with $V/t$ for small $V/U$. This
finding is of great significance to the cold-atom experiments,
where an increased Neel temperature would lead to an easier detection
of the AF ordered state as the current OLE experiments on the
  standard fermionic Hubbard model are having difficulties reaching
  the $T<T^N$ regime.

\begin{figure}[t]
\includegraphics[width=\columnwidth]{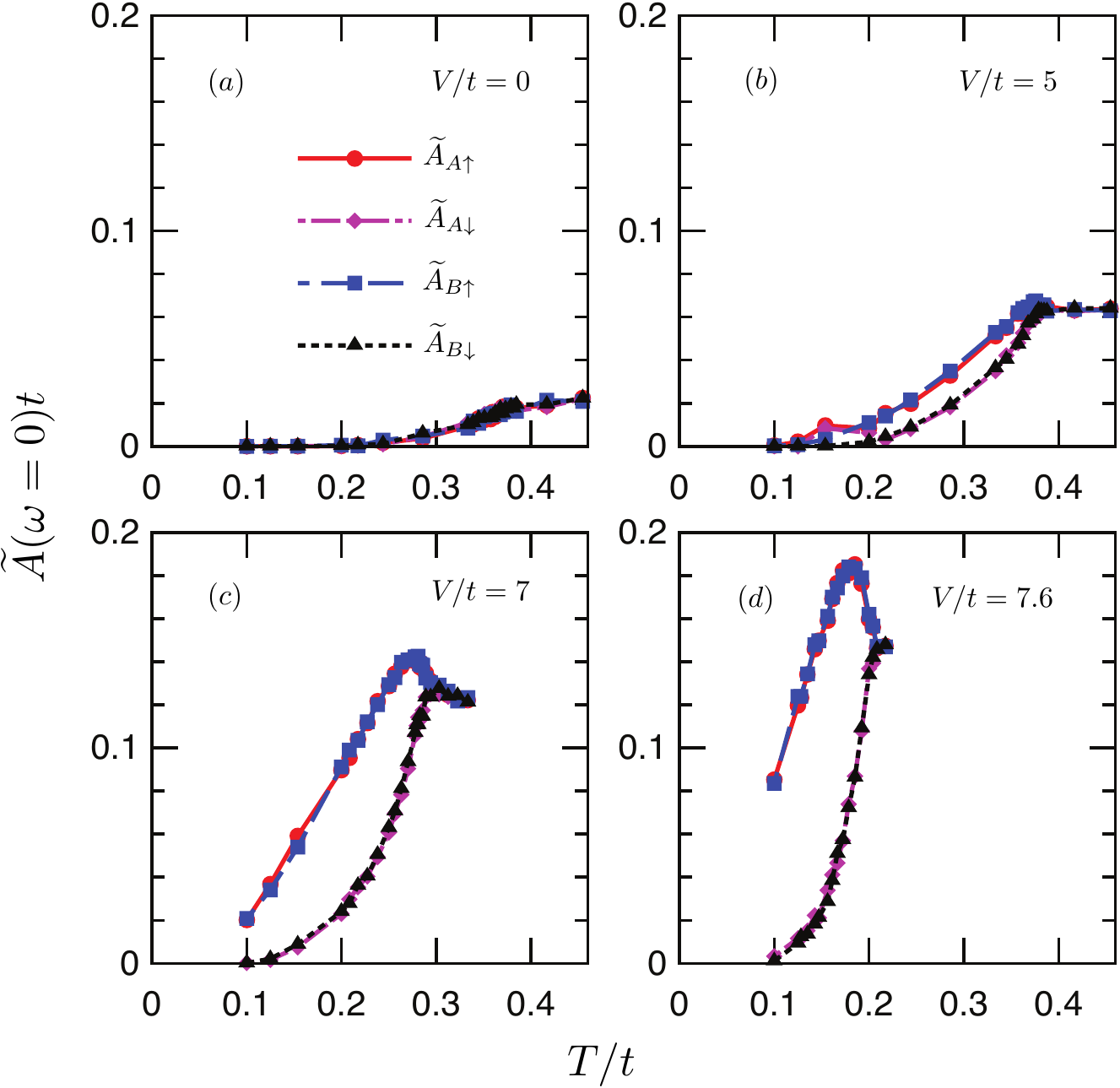}
\caption{Estimated zero frequency single particle density of states
  $\widetilde{A}(\omega=0)$ of
  $\up$ and $\dn$ spin particles on $A$ and $B$  sublattices,
 as a function of temperature 
for (a) $V/t=0$,
 (b) $V/t=5.0$, (c) $V/t=7.0$ and (d) $V/t=7.6$. The Hubbard
 interaction is fixed at $U/t=10$ for all the plots. As $V/t$
 increases, the $\up$ spin
DOS show a marked
increase over the $\dn$ spin DOS.}
\label{fig:3}
\end{figure}

The perturbative argument, which predicts a divergent super-exchange  coupling at $V=U$, 
breaks down as $V/U$ approaches unity. The DMFT results for $T^N$ as a
function of $V/t$ (see Fig~\ref{fig:2}), however, show that for the
strongly interacting system, $T^N$ continues its rise upto $V/U \sim
0.6$, and the optimum $T^N$ is $40\%$ higher than that of the
standard Hubbard model for $U/t=16$.

In the ionic Hubbard model,
when $V/U \sim 1$, the
potential energy gained by fermions on $A$ sublattice can compensate for the energy cost of
forming a double occupancy (as long as it is formed on the $A$
sublattice). Thus the state {$|\!\!\up_A,\dn_B\rangle$} and the state
{$|\!\!\up_A\dn_A,0_B\rangle$}, will have similar energy differing by $\sim
U-V$. For $U-V\sim J$, these states lie in the low
energy subspace of the system and the Hubbard model can no longer be
reduced to a simple Heisenberg model even at half-filling. This
picture is supported by the fact that the sublattice density asymmetry, $\Delta n=n^\up_A+n^\dn_A-n^\up_B-n^\dn_B$, remains
constant upto $T^N$ for small $V/U$,
while for $V/U\sim 1$, it rapidly rises with temperature from a low
temperature asymptotic value to a constant high temperature value
beyond the Neel transition. This is clearly seen in
Fig.~\ref{fig:1}(b). At half-filling, $\Delta n$ is a
measure of the relative weight of doubly occupied states in
the thermal ensemble, with $\Delta n=0$ for states in the projected
subspace, while $\Delta n=2$ for the state with
perfect density order, double occupancies on sublattice $A$ and
vacancies on sublattice $B$. The temperature dependence of
the density asymmetry shows that at small $V/U$, configurations with
double occupancies do not play a major role in spin disordering, while
at large $V/U$ the loss of AF order is driven by increasing presence
of such configurations.

\begin{figure}[t]
\includegraphics[width=\columnwidth]{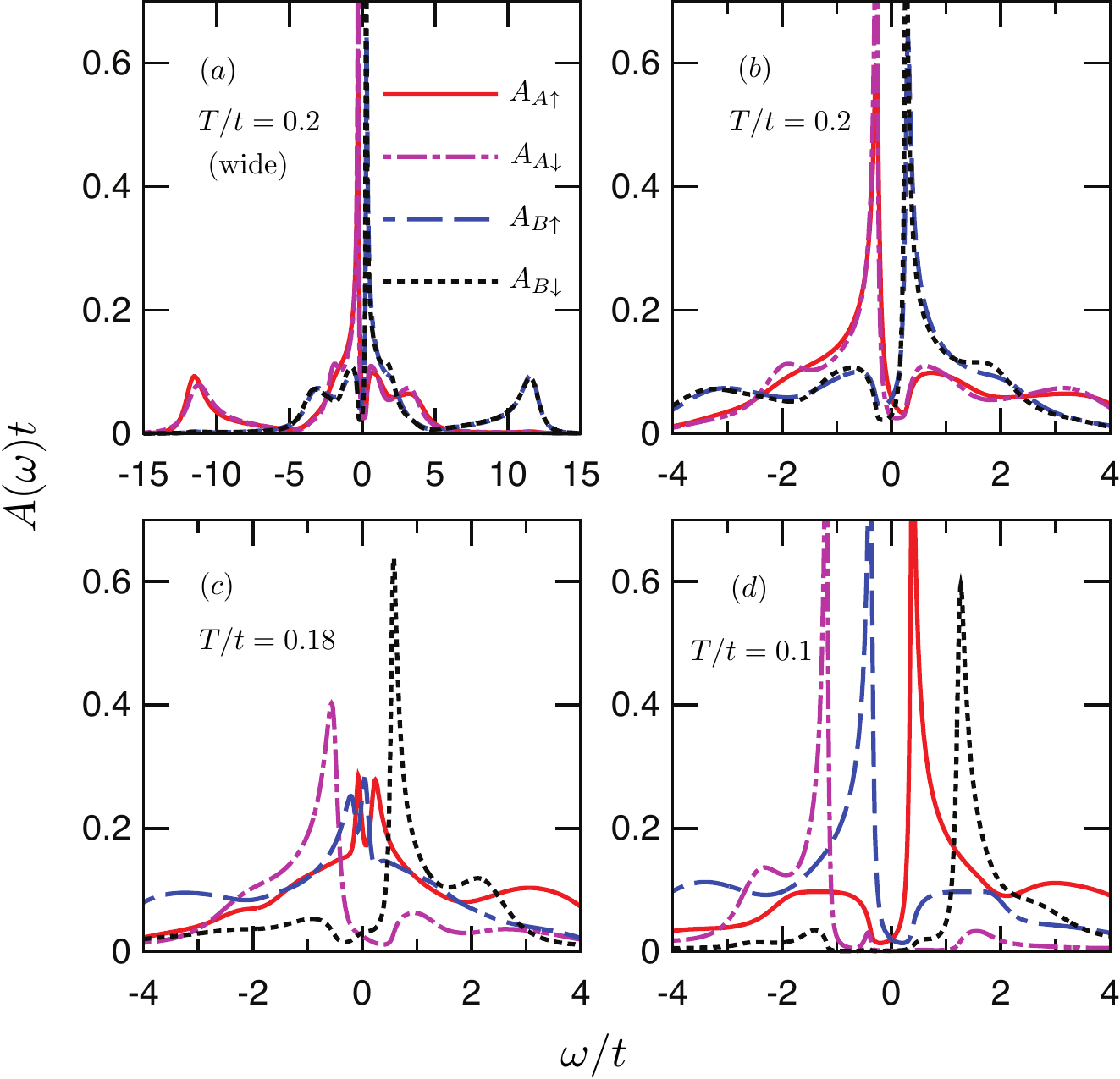}
\caption{The DOS of $\up$ and $\dn$ spins on $A$ and $B$
sublattices, calculated by maximum entropy method, as a function of
energy for $V/t=7$ and $U/t=10$. (a) The results
for $T/t=0.2$ ; Low energy results for same temperature is shown in (b). (c) and (d) show the low energy results for
$T/t=0.18$ and $T/t=0.1$ respectively.}
\label{fig:4}
\end{figure}
The loss of AF ordering due to inclusion of double occupancies results in an 
apparently counter-intuitive phenomenon. As $V$ is increased close
to $U$, the low energy single particle DOS shows spin-polarization,
which initially increase with increasing temperature, before vanishing
at the AF transition point. Thus, the system
exhibits static AF order, but the low energy dynamics
of the system is very similar to a ferromagnet. Specifically, if a
current is set up in the system, it will be carried by spin-polarized
carriers and the system would show a finite spin-conductivity even in the
absence of a magnetic field. To see this, we compute the zero frequency DOS for the fermions on a
given sublattice and with a given spin from the imaginary-time Green's functions through 
\beq
\widetilde{A}_{\sigma,\alpha}(\omega=0)= -\frac{1}{\pi T}G^\sigma_\alpha(\tau=1/(2T)).
\label{eq:zfdos}
\eeq 
 This
approximation to the zero frequency density of states has been
frequently used to determine metallic vs. insulating nature
of the system \cite{Trivedi95}. In Fig.~\ref{fig:3}, we plot the zero
energy density of states for fermions of both spins on both sublattices as a function of temperature for (a) $V/t=0$, (b)
$V/t=5$, (c) $V/t=7$ and (d) $V/t=7.6$ for a system with
$U/t=10$. Upto $V/t=5$, we see very little spin asymmetry in the DOS,
while for $V/t$ larger than $7$, there is a
large asymmetry in the zero energy DOS of the $\up$ and $\dn$
spins. This asymmetry grows with temperature, and reaches a peak close
to the transition before vanishing at the transition. Thus, the
system is metallic at these temperatures, with transport and low energy
dynamics dominated by $\up$ spins.  


To understand the mechanism of spin-polarization of low energy DOS, we consider
the extreme case of  a perfectly Neel ordered state with $\dn$ spins
on $A$ sublattice and $\up$ spins on $B$ sublattice. As $V/U$ is
increased close to $1$, the $\up$ spin on the $B$ sublattice can move to
$A$ sublattice to form a double occupancy and keep the state in the
low energy subspace.  However, the $\dn$ spin on the $A$ sublattice
cannot move to the $B$ sublattice, as the resulting state would have a
high energy $\sim U+V$. These processes are thus prohibited to ${\cal
  O}(t)$ and can only happen with a scale ${\cal O}[t^2/(U+V)]$. Thus
the low energy dynamics is mainly the dynamics of the $\up$ spins in
this case. Once this basic mechanism is understood, one can generalize
to more complicated states, but as long as the AF ordering is present,
the low energy density of states would be dominated by the majority
spins on the $B$ sublattice. The spin-polarization of the low energy
density of states is thus understood as a consequence of a resonance
between a projected AF
ordered state and a state with double occupancy on the $A$
sublattice. This shows the role of double occupancies in the degradation of AF ordering in the
ionic Hubbard model for $V/U\sim 1$.

In order to study the energy dependence of the
spin asymmetry in the DOS, we
analytically continue the Matsubara Green's functions to the real
frequency domain and obtain the frequency dependent spectral weight
(spin and sublattice resolved), using the
method of Ref.~\cite{Wang09}. The results are plotted in
Fig.~\ref{fig:4} for a system with $U/t=10$ and $V/t=7$ for three
different temperatures. In Fig.~\ref{fig:4}(b) we plot the results at
a temperature of $T/t=0.2$, which is above the AF transition
temperature.  Each
sublattice shows complete symmetry between $\up$ and $\dn $ spins
in terms of the spectral weight.
The
asymmetry between $A$ and $B$ sublattices reflects the different
densities on these sublattices. In Fig.~\ref{fig:4}(c), we plot the
results for a temperature $T/t=0.18$, which is just below the
transition temperature, and where the spin asymmetry of the zero
frequency DOS, as seen in Fig.~\ref{fig:3}(c), is maximal. In this case,
we clearly see a buildup of low energy DOS for the $\up$ spin, while
the $\dn$ spin DOS shows a soft gap. Finally in Fig.~\ref{fig:4}(d), we
plot the low temperature results at $T/t=0.1$. In this case, the $\up$
spin DOS shows a soft gap, while the $\dn$ spin DOS shows a hard gap
in the spectrum.
 
In conclusion, we have studied the ionic fermionic Hubbard model using DMFT. This model can be easily
implemented in cold atom optical lattice systems and has a higher Neel
temperature for AF transition than the standard Hubbard model for an
accessible region in the parameter space. For small $V/U$,
the effective super-exchange scale, given by
$4t^2/(U-V^2/U)$, increases with the ionic potential $V$. As a
consequence, $T^N$ increases with $V/U$, reaches an optimum value
around $V/U\sim 0.6$ and then goes down with further increase in
$V/U$. The optimum temperature is about $40\%$ higher than that of the
standard Hubbard model for $U/t=16$, which should help in observing
super-exchange dominated AF ordering in OLE experiments.  At large
$V/U\sim 1$, the AF order is degraded by inclusion of more and more
configurations with double occupancies (on the $A$ sublattice) in the
ensemble. A consequence of this mechanism is the surprising result
that the low energy density of states shows strong spin asymmetry in
the AF phase.  Thus, with respect to dynamics and transport, the
system behaves like a ferromagnet, although it shows static AF spin
ordering. This novel feature of the ionic Hubbard model should manifest
itself in the optical lattice emulation experiments.

\begin{acknowledgments}

The authors thank M. Randeria, H. R. Krishnamurthy and K. Sengupta for
fruitful discussions. This work is supported by  the NSF-JQI-PFC, AFOSR MURI and ARO MURI.
The DMFT code is developed based on the ALPS library \cite{ALPS}. The
computations are conducted at the Center for Nanophase Materials
Sciences, which is sponsored at Oak Ridge National Laboratory by the
Scientific User Facilities Division, Office of Basic Energy Sciences,
U.S. Department of Energy.
\end{acknowledgments}

\end{document}